\documentclass[pra,aps,amsmath,amssymb,superscriptaddress,twocolumn,longbibliography]{revtex4-1}

\usepackage{graphicx,multirow}
\usepackage{color,outlines}
\usepackage[english]{babel}
\usepackage{amsthm}
\usepackage{hyperref}
\usepackage{physics}
\usepackage{outlines}

\usepackage{tabularx}
\usepackage{array}
\usepackage{float}
\usepackage{tikz}
\usepackage{adjustbox}
\usetikzlibrary{positioning}

\usepackage[sc,osf]{mathpazo}
\newcolumntype{Y}{>{\centering\arraybackslash}X}

\newtheorem{defn}{Definition}

\definecolor{green2}{RGB}{0,100,0}

\begin{document}

\title{Four Party Absolutely Maximal Contextual Correlations}

\author{Nripendra Majumdar}
\email{nripendrajoin123@gmail.com}
\affiliation{Department of Physics,\\
Indian Institute of Technology Tirupati, Tirupati, India~517619}

\begin{abstract}
The Kochen-Specker theorem revealed contextuality as a fundamental non-classical feature of nature. Non-locality arises as a special case of contextuality, where entangled states shared by space-like separated parties exhibit nonlocal correlations. The notion of maximality in correlations, analogous to maximal entanglement, is less explored in multipartite systems. In our work, we have defined maximal correlations in terms of contextual models, which are analogous to absolutely maximally entangled (AME) states. Employing the sheaf-theoretic framework, we introduce maximal contextual correlations associated with the corresponding maximal contextual model. The formalism introduces the contextual fraction $\mathrm{CF}$ as a measure of contextuality, taking values from $0$ (non-contextual) to $1$ (fully contextual). This enables the formulation of a new class of correlations termed absolutely maximal contextual correlations (AMCC), which are both maximally contextual and maximal marginals.  In the bipartite setting, the canonical example is the Popescu–Rohrlich (PR) box, while in the tripartite case, it includes Greenberger–Horne–Zeilinger (GHZ) correlations and three-way nonlocal correlations. In this work, we extend these findings to four-party correlations. Notably, no AME state exists for four qubits, which introduces a subtle difference between AMCC and AME. The construction follows the constraint satisfaction problem (CSP) and parity-check methods. In particular, the explicit realization of a non-AMCC correlation that is maximally contextual yet not maximal marginal is obtained within the CSP framework.
\end{abstract}

\maketitle

\section{Introduction}
Contextuality is a nonclassical feature that defies the consistent valuation of measurable quantities, independent of what other quantities are measured alongside it. The initial development, followed by the KS theorem \cite{KStheorem1990}. This theorem provides a foundational basis for understanding nature through logical proofs. Their original construction employed 117 vectors \cite{KStheorem1990}, which was technically complex. However, later simpler proofs, such as the Mermin square \cite{MerminSquare1990}, 18-ray vectors \cite{Cabello18vectors1996}, and the 13-ray vector proof \cite{KS13rayproof}, known as state-independent proofs, provided a profound understanding of contextuality. The state-dependent proof on a qutrit \cite{KCBS2008} shows that contextuality can be exhibited in a single system and does not require spatial separation. This requires a contextual inequality analogous to the CHSH inequality. A deeper understanding of contextuality can be seen through the sheaf-theoretic notion of contextuality \cite{Sheaftheory2011}. This provides a general mathematical tool that yields a measure of contextuality, the Contextual Fraction $\mathrm{CF}$ \cite{CFPhysRevLett.119.050504}, which lies in $[0,1]$. Sheaf theory addresses the inconsistency in constructing global data from local data. This aligns with the existing results of Fine's theorem \cite{Finetheorem}. In this regard, it is evident that non-locality is a special case of contextuality. The well-known example is the Bell-type model $(n, m, o)$, which consists of $n$ parties, each assigned $m$ measurements, and each measurement has $o$ outcomes. Under the sheaf-theoretic notion, such models are inscribed as measurement scenarios. The joint measurements are defined as the set of compatible measurements (context), called the measurement cover. Considering the measurement cover, the set of outcomes for each joint measurement, along with the probability distribution over these outcomes, defines the empirical model. If any empirical model satisfies the no-signaling condition, then it is a no-signaling empirical model. The mentioned mathematical model treats any empirical model as a linear system; if such a system has no non-negative solution, then the associated empirical model is contextual; otherwise, it is non-contextual. Any empirical model can be decomposed into a non-contextual model and a contextual model, where the coefficient corresponding to the non-contextual model is the non-contextual fraction $\mathrm{NCF}=p$, and for the contextual model, it is the contextual fraction $\mathrm{CF}=1-p$. Any empirical model has three forms: if $\mathrm{CF}=0 (p=1)$, then the model is non-contextual, and $\mathrm{CF=1}(p=0)$ for the maximally contextual model; otherwise, the model is contextual for $0<\mathrm{CF}<1$ \cite{CFPhysRevLett.119.050504}. Any empirical model is strongly contextual iff it is maximally contextual \cite{Sheaftheory2011}.

The maximal contextual model is of importance. It demonstrates the computational advantage, such as upgrading a classical computer's power from linear to non-linear function, and higher polynomial function evaluation when the underlying resources are maximally contextual; such examples are studied in Measurement-Based Quantum Computation (MBQC) \cite{MBQCAnders2009, CFPhysRevLett.119.050504, CtxinMBQC2013, MBQCbeyondqubit2018} models. Contextuality has been studied in other mathematical frameworks as well, such as cohomology, which exhibits global obstructions \cite{Cohominctx, CohomAbramskyMansfieldBarbosa2011, HomotopyRaussendorf2020, Peresconjec2020, FRWalleghem2024, FinitedimMBQCRaussendorf2023, SahaHorodeckiPawlowski2019}. The concept of global obstruction can be prominently understood in bundle diagrams \cite{Bundlediagram2018}.

Recent advancements provide experimental evidence for contextual models such as Mermin's square and the KCBS inequality \cite{Merminnature2009, KCBSnature2017}.
Maximal contextuality has been studied in many-body systems to understand the phase, and the quantification of contextuality has been examined to comprehend the quantum advantage over classical counterparts in non-local games \cite{Nonlocalgame2025, Manybodynoncontextual2025, ContextualityAI2025}. The experimental verification of contextuality in randomness certification is also being studied \cite{Randomness2025}.

Since Non-locality is a special case of contextuality, the correlations associated with any contextual model are termed contextual correlations. However, the correlations corresponding to maximal contextual models are called maximal contextual correlations. Contextual correlations (in all forms) cannot be generated by any non-contextual hidden-variable (NCHV) model, which is closely related to the notion of realism discussed in the EPR paper \cite{EPR1935}. The long-standing debate eventually culminated in Bell's Inequality \cite{Bell1964}, which provides bounds on local theories and characterizes nonlocal correlations. Such correlations arise from particular quantum states, which Schrodinger later identified as exhibiting entanglement. Entanglement theory is rich in several aspects and plays a crucial role in quantum information theory. The classification and quantification of entanglement have been accomplished within the framework of local operations and classical communication (LOCC) \cite{QuantumentglHodric2009}, and a unique measure of entanglement has been defined. In a bipartite system, there is only one class of maximally entangled states (Bell states), and von Neumann entropy is the only measure of entanglement \cite{ThermoandEntanglement, Conditionofentanglement, Threequbitclass}. For a three-qubit system, there are two inequivalent classes of genuine tripartite entanglement (the GHZ class and the W class) under stochastic LOCC (SLOCC)\cite{Threequbitclass}. However, for general multipartite systems, classifying and quantifying entanglement becomes intractable. Despite substantial progress \cite{Fourqubitclass, LUeuivalentclassKraus2010, MaxEntmultiparty2013, Classofentmutlidim, NessandsuffconditionLM2019, EntanglementLO2017, EntanglementtransformLi2024, EntanglementtranformApprox, Multipartymeasureentang, Trianglemeasure}, this question remains open for multipartite systems: does a notion of a ``maximally entangled state'' exist, and if so, how should it be defined?

The characterization of entangled states and correlations is closely related: entangled states are not factorizable into subsystem states, and this non-separability can manifest as non-factorizable correlations. However, the relationship is not one-to-one, so the theories of entanglement and nonlocal correlations are deeply intertwined and conceptually subtle \cite{InterplaynonlocalEnatnglement2023}. This perspective is evident within the framework of local operations and shared randomness (LOSR) \cite{TypeIndependentSchmid2020, TypeSpaceLOSR}, which serves as the analog of LOCC in entanglement theory. In the multipartite case, the more general framework is provided by wiring and classical communication prior to inputs (WCCPI) \cite{OperationalNonlocality, OperationalMultiparty2020}, which reveals that the earlier definitions \cite{DefinitionnonlocalityGisin} are inadequate and require refinement. Even after identifying the ‘golden unit’ of entanglement in the bipartite case and the maximally entangled classes in the three-qubit system under LOCC and SLOCC, the concepts of entanglement distillation and concentration face limitations when one attempts to map entanglement theory to correlations. Consequently, maximal entanglement does not necessarily correspond to maximal correlations. As a result, in the asymptotic limit, many sets of correlations are incomparable \cite{AsymptoticQuantumNonlocality}. The rigorous formulation of maximal entanglement and maximal correlations in genuine multipartite systems remains an open and actively investigated problem.

To extend this understanding of maximality in multiparty systems, we have considered a special class of entangled states, namely absolutely maximally entangled (AME) states \cite{AME}. These pure states are maximally entangled and maximally mixed in their reduced bipartition. We defined an analogous class of correlations, termed absolutely maximal contextual correlations (AMCC) \cite{AMCC}, which are maximal contextual correlations whose reduced correlations are maximally random (i.e., maximal marginals). These maximal correlations do not contain any local fraction \cite{GeometriQuantumSetAshutosh, AVN}. In the $(2,2,2)$ scenario, the AMCCs are the $8$ PR boxes, none of which admits a quantum realization. In the two-party qubit case, the AME states are Bell states; however, the associated correlations are not maximal. In contrast, the AMCCs in the $(3,2,2)$ scenario are GHZ correlations, which arise from the GHZ state, an AME state in a three-party qubit system. Beyond this particular correlation in the $(3,2,2)$ scenario, there exist infinitely many such correlations lying on the face of the polytope. In this work, we present extended results beyond the $(3,2,2)$ scenario. While no AME states exist in the four-qubit case, we find that infinitely many AMCCs exist in the $(4,2,2)$ scenario. AMCC construction is carried out using the parity-check method and the constraint satisfaction problem (CSP) framework. Furthermore, apart from AMCCs, there exists another class of correlations that are maximally contextual but not maximally random; we refer to them as non-AMCCs.

Here is the outline for the rest of the paper. In Section~(\ref{sec 00}), we introduce the mathematical formalism for correlations and contextual correlations. Section~(\ref{sec 01}) defines the absolutely maximal contextual correlation. In Section~(\ref{sec 02}), we present the construction of AMCC and non-AMCC. Finally, Section~(\ref{sec: 03}) follows the conclusion.

\section{Formalism}
\label{sec 00}
Correlations arising from measurements on entangled states exhibit nonclassical features and can serve as valuable resources. A deeper understanding can be obtained in a Bell-type scenario, where the entire setup can be represented as a ``black-box'' framework. This abstraction makes the correlations more powerful and can also be used in device-independent protocols \cite{RNG2010, DISecretsharing}. Non-factorizable, nonlocal correlations violate Bell’s inequality and form a larger set than those correlations that satisfy it (local correlations) \cite{GeometriQuantumSetAshutosh}. Quantum mechanically, the maximum violation achievable is $2\sqrt{2}$ (Tsirelson’s bound) \cite{Tsirelson1980}, whereas PR boxes—being post-quantum correlations—reach the algebraic maximum value of $4$ \cite{PRBox1994}. These are the no-signaling correlations, which form a superset that includes both quantum and local correlations. For illustration, consider a two-party system. Alice and Bob each possess a box with two input choices, $\{0,1\}$, corresponding to two measurement settings, $\{Y_1, Y_1^\prime\}$ and $\{Y_2, Y_2^\prime\}$, respectively. The correlations are described by the set of joint conditional probabilities $q(y_1, y_2 \mid Y_1, Y_2)$, which specify the probability of obtaining outcomes $y_1$ and $y_2$ given that Alice and Bob choose measurement settings $Y_1$ and $Y_2$, respectively. This represents a $(2,2,2)$ scenario, where $m=2$, $n=2$ and $o=2$.
For the general setting $(n,m,o)$, the correlations \( \{q(y_1,\dots,y_n \mid Y_1,\dots,Y_n)\}\) satisfy the following properties: they are normalized \(\sum_{y_1,\dots,y_n} q(y_1,\dots,y_n \mid Y_1,\dots, Y_n) = 1\) for all \(Y_1,\dots,Y_n\); positive  \(q(y_1,\dots,y_n \mid Y_1,\dots,Y_n) \geq 0\) for all outcomes and settings; and satisfy the no-signaling condition, namely that the marginal probability is independent of the measurement choices corresponding to $k$ and $k\prime$ parties:
\begin{widetext}
\begin{equation}
        \sum_{y_k}q(y_1,\cdots,y_k,\cdots,y_n\mid Y_1,\cdots,Y_k,\cdots,Y_n)=
\sum_{y_k}q(y_1,\cdots,y_k,\cdots,y_n \mid Y_1,\cdots,Y_{k\prime},\cdots,Y_n),
\label{eq: no-signaling condition}
\end{equation}
\end{widetext}
for all $k$, all outcomes $y_1,\cdots,y_{k-1},y_{k+1},\cdots,y_n$, and all settings $X_1,\cdots,X_{k-1},X_{k+1},\cdots,X_n$, where the index $k\in \{1,\cdots,n\}$, each party $k$ has $m_k$ measurement choices $X_k\in \{1,\cdots,m_k\}$, and for each measurement choice $X_k=j$ of party $k$, there are $o_{kj}$ possible outcomes $y_k\in \{1,\cdots,o_{kj}\}$. The set of correlations $P$ satisfying the no-signaling condition constitutes a convex polytope, referred to as the no-signaling polytope \cite{LiftingBellIneq2005, GeometriQuantumSetAshutosh}.
The correlations are said to be \emph{local} if they are reproducible by any local hidden-variable model, namely
\begin{widetext}
\begin{equation}
    q(y_1,\dots,y_n\mid Y_1,\dots,Y_n)= \int d\mu\,\zeta(\mu)\,q(y_1\mid Y_1, \mu)\cdots q(y_n\mid Y_n,\mu),
\end{equation}
\end{widetext}
where $\zeta(\mu)\geq 0$, $\int d\mu\,\zeta(\mu)=1$, and $q(y_i\mid Y_i, \mu)$ denote the marginal probability of obtaining the outcome $y_i$ given the measurement choice $Y_i$ and the hidden variable $\mu$.
The \emph{local polytope} \cite{LiftingBellIneq2005} consists of the set of local correlations, which is the convex hull of deterministic points. Any Bell inequality $\mathbf{b}\cdot\mathbf{q}\geq b_0$, where $(\mathbf{b},b_0)\in \mathbb{R}^{h+1}$, associated with the Bell polytope \cite{LiftingBellIneq2005} $P^B\subseteq \mathbb{R}^h$, is said to be valid if it is satisfied by all $\mathbf{q}\in P^B$; equivalently, it suffices that it is satisfied by all extreme points of $P^B$. We are particularly interested in correlations lying on a face of the no-signaling polytope $F^{NS} = \{\mathbf{q} \in \mathbf{NS}\;|\; q = 0\}$ \cite{AVN, AMCC}, where $q$ denotes the specific component of the vector $\mathbf{q} = \{q(y_1, \cdots, y_i \mid Y_1, \cdots, Y_i)_{i \leq \dim P - 1}\}$, and $\mathbf{NS} \subset \mathbb{R}^h$ represents the set of no-signaling correlations. These correlations are expressed in a maximally reduced form via zero constraints and lie on a lowest-dimensional face of the no-signaling polytope \cite{AVN, AMCC}. Moreover, such correlations satisfy the all-versus-nothing (AVN) argument and consequently possess no local fraction. We refer to these correlations as maximal correlations, which are discussed in the next section.

\subsection{ Contextual correlations}
\label{subsec 00}
In this section, we formalize the general experimental scenario. Any dataset obtained from an experiment can be inscribed as $(n,m,o)$-scenario, represented by a triplet $\langle \mathcal{Y}, \mathcal{M}, \mathcal{O} \rangle$, where $\mathcal{Y}$ is a set of all observables, $\mathcal{M}$ is a measurement cover, and $\mathcal{O}$ is a set of possible outcomes associated with each measurement. These concepts are formalized in the definitions that follow. The accompanying examples are presented in the context of the $(4,2,2)$-scenario.

\begin{defn}[Measurement Cover] 
    The measurement cover is a set $\mathcal{M}\subseteq \mathcal{Y}$ such that:
    \begin{enumerate}
        \item $\bigcup \mathcal{M}=\mathcal{Y}$
        \item The chain condition: if $\mathcal{C},\mathcal{C}^\prime\in \mathcal{M}$ and $\mathcal{C}\subseteq\mathcal{C}^\prime$, then $\mathcal{C}=\mathcal{C}^\prime$.
    \end{enumerate}
\end{defn}
For illustration, the measurement set for $(4,2,2)$-scenario,
\begin{equation}
    \mathcal{Y}=\{Y_1,\ Y_1^\prime,\ Y_2,\ Y_2^\prime,\ Y_3,\ Y_3^\prime,\ Y_4,\ Y_4^\prime\},
    \label{422 measurment set}
\end{equation}
the measurement cover consists of $16$ contexts (see section ~(\ref{subsec 05}))
and the outcome set is $\mathcal{O}=\{0,1\}$. The joint outcome associated with each context is named section $s$.

\begin{defn}[section]
    A section $\mathrm{s}$ is a mapping from $\mathcal{U}\subseteq \mathcal{M}\subseteq\mathcal{Y}$ to $\mathcal{O}$, i.e.,
    \begin{equation}
        \mathrm{s}:\mathcal{U}\to\mathcal{O}.
    \end{equation}
\end{defn}
The \emph{presheaf} (seaf of events) $\mathcal{E}: P(Y)^{\text{op}} \to \mathsf{Set}$, defines the set consisting of all the sections, which is a functor \cite{Sheaftheory2011}. Namely, in the previous example, for each context $\mathcal{C}\in \mathcal{M}$, $16$ sections $(\mathcal{E}(U)=\{(0,0,0,0),(0,0,0,1),\cdots,(1,1,1,0),(1,1,1,1)\})$ are possible.
The composition of the distribution function $\mathcal{D}_{\mathcal{R}}\mathcal{E}:P(Y)\to \mathsf{Set}$ with the presheaf provides the weight assignments over the set of joint outcomes defined by the sheaf of events, where $\mathcal{D}_{\mathcal{R}}:\mathsf{Set}\to \mathsf{Set}$ is the distribution functor that sends a set $\mathcal{E}$ to
\[
\mathcal{D}_{\mathcal{R}}(\mathcal{E}) = \{\mathcal{P} : \mathcal{E} \to \mathbb{R}_{\geq 0} \mid \sum_{y \in \mathcal{E}} \mathcal{P}(Y) = 1\},
\] and $\mathcal{R} = \{\mathbb{R}_{\geq 0}, +, \cdot, 0, 1\}$ is a semiring of non‑negative reals. This weight assignment is the presheaf of probability distribution. Given the distribution $\mathrm{e} \in \mathcal{D}_{\mathcal{R}}(\mathcal{E}(U^\prime))$ over sections corresponding to $U^\prime\subseteq \mathcal{Y}$, there exists a restriction map \(\text{res}_U^{U^\prime}\) for distributions such that
\begin{equation}
    \begin{split}
        \mathcal{D}_\mathcal{R}(\mathcal{E}(U^\prime)) &\xrightarrow{\text{res}_U^{U^\prime}} \mathcal{D}_{\mathcal{R}}(\mathcal{E}(U)),\\
        \mathrm{e} &\mapsto \mathrm{e}|_U.
    \end{split}
    \label{eq:restriction-map}
\end{equation}
Here $\mathrm{e}|_U$ is the marginal over all $\mathrm{s} \in \mathcal{E}(U)$:
\begin{equation}
    \mathrm{e}|_U(\mathrm{s}) = \sum_{\substack{\mathrm{s}^\prime \in \mathcal{E}(U^\prime) \\ \mathrm{s}^\prime|_U = \mathrm{s}}} \mathrm{e}(\mathrm{s}^\prime).
    \label{eq:marginal-sum}
\end{equation}

\begin{defn}[Empirical model]
    Given a measurement scenario $\langle \mathcal{Y}, \mathcal{M}, \mathcal{O} \rangle$, an \emph{empirical model} is a family of distributions $\{\mathrm{e}_{\mathcal{C}}\}_{\mathcal{C}\in\mathcal{M}}$, where each $\mathrm{e}_{\mathcal{C}}\in \mathcal{D}_{\mathcal{R}}(\mathcal{E}(\mathcal{C}))$.
\end{defn}

\begin{defn}[Generalized no‑signaling] 
    Let $\mathcal{C}_1, \mathcal{C}_2\in \mathcal{M}$ with $\mathrm{e}_{\mathcal{C}_1}\in \mathcal{D}_\mathcal{R}\mathcal{E}(\mathcal{C}_1)$ and $\mathrm{e}_{\mathcal{C}_2}\in \mathcal{D}_\mathcal{R}\mathcal{E}(\mathcal{C}_2)$. The generalized no‑signaling condition is
    \begin{equation}
        \mathrm{e}_{\mathcal{C}_1}|_{\mathcal{C}_1\cap \mathcal{C}_2}=\mathrm{e}_{\mathcal{C}_2}|_{\mathcal{C}_1\cap \mathcal{C}_2}.
        \label{eq:generalized-nosignal}
    \end{equation}
    An empirical model satisfying this compatibility condition is called a \emph{no‑signaling empirical model}.
\end{defn}

The set of all the sections over $\mathcal{Y}$, denoted by $\mathcal{E}(\mathcal{Y}) = \mathcal{O}^{\mathcal{Y}}$, defined by the map $g:\mathcal{Y}\to \mathcal{O}$, is referred to as the \emph{global sections}. The distribution over these sections $\mathrm{e}_\mathcal{Y} \in \mathcal{D}_\mathcal{R} \mathcal{E}(\mathcal{Y})$, are called the \emph{global distribution}, and it also satisfies marginality condition
\begin{equation}
    \mathrm{e}_\mathcal{C}(\mathrm{s}_\mathcal{C}) = \sum_{\substack{\mathrm{s}^\prime \in \mathcal{E}(\mathcal{Y}) \\ \mathrm{s}^\prime|_\mathcal{C} = \mathrm{s}_\mathcal{C}}} \mathrm{e}_\mathcal{Y}(\mathrm{s}^\prime).
    \label{eq:global-marginal}
\end{equation}
Hence, ``if the model admits a global distribution, then it can be described by a deterministic hidden-variable model.''
The question of whether such a global distribution exists can be formulated as a linear-algebraic feasibility problem. Given a measurement scenario $\langle \mathcal{Y}, \mathcal{M}, \mathcal{O} \rangle$ with $n_1$ global sections and $n_2$ local sections, the restriction map is determined by the incidence matrix  $[M]_{n_1\times n_2}$. Provided an empirical model, if the global distribution exists $\mathbf{d}\in \mathcal{D}_\mathcal{R} \mathcal{E}(\mathcal{Y})$, then the following associated linear system must have non-negative solutions over the positive semiring:
\begin{equation}
    \mathbf{M}\,\mathbf{d} = \mathbf{v}, \qquad \mathbf{d} \geq \mathbf{0},
    \label{eq:linear-system}
\end{equation}
where $\mathbf{M}$ is the incidence matrix, and $\mathbf{v}$ is the vector of the probability distribution over local sections.
\begin{defn}[Contextual]
    A no‑signaling empirical model $\{\mathrm{e}_{\mathcal{C}}\}_{\mathcal{C}\in \mathcal{M}}$ is \emph{contextual} if and only if the linear system in Eq.~(\ref{eq:linear-system}) admits no solution over the non‑negative reals.
\end{defn}
A possibilistic empirical model is defined over a Boolean semiring instead of a semiring of non-negative reals.

\begin{defn}[Strong contextuality]
    Given a possibilistic model derived from an empirical model, if there exists no global section $g$ such that $\forall \mathcal{C}\in\mathcal{M},\; \mathrm{e}_{\mathcal{C}}(g|_{\mathcal{C}}) > 0$, then the model is called \emph{strongly contextual}.
    \label{def: Strong contextuality}
\end{defn}
Any empirical model $M$ can be expressed in the following convex decomposition form:
\begin{equation}
    M=pM^{\mathrm{NC}}+(1-p)M^{\mathrm{SC}},
    \label{eq:convex-decomp}
\end{equation}
where the contextual fraction is defined as $\mathrm{CF}=1-p$ and the non-contextual fraction as $\mathrm{NCF}=p$. The condition $p=0$ corresponds to ``maximally contextual'', whereas $p=1$ denotes a noncontextual model. Furthermore, a model is strongly contextual iff it is maximally contextual. The simplex method is a linear optimization technique that can be implemented to compute the contextual fraction $\mathrm{CF}$. This optimization problem consists of an objective function and linear constraints that collectively define a multidimensional convex polytope. The simplex method consists of the following three ingredients: slack variables, a tableau, and pivot operations. This procedure provides an efficient algorithm to compute $\mathrm{CF}$.

Since the maximality of correlations corresponding to the maximality of quantum states in a multiparty system remains unexplored, we have used AME states as a reference for an analogous understanding of correlations, since these states are maximally mixed in their bipartite reduced form. Accordingly, we have defined a new class of correlations that are maximally contextual (correlations associated with maximal contextual models) and maximal marginals, presented in the next section. 

\section{Absolutely maximal contextual correlations}
\label{sec 01}
The contextual fraction $\mathrm{CF}$, provides a quantitative measure of contextuality. The $\mathrm{CF}=1$ is for maximal contextual models, and other forms are weaker than this. The mathematical framework discussed in the section~(\ref{sec 00}) is employed to define a new class of correlations: maximal contextual correlation with maximal marginal, called absolutely maximal contextual correlations (AMCC) \cite{AMCC}. Since maximally contextual empirical models exhibit all-versus-nothing (AVN) proofs, the associated correlations lie on a face of the no-signaling polytope (FNS) correlations and contain no local fraction \cite{AVN, AMCC}. The formal definitions are provided below.
\begin{defn}[Maximal Marginals]
For any $(n,m,o)$ Bell scenario, the correlations are called \emph{maximal marginals} if all marginals satisfy
\begin{equation}
    e\mid_U = p(y_1,\dots,y_k \mid Y_1,\dots,Y_k) = \frac{1}{2^k} \quad \forall \; k < n,
\end{equation}
where $U = \{Y_1,\dots,Y_k\} \subset \mathcal{Y}$ and each $y_i \in \{0,1\}$.
\label{def:max-mixed}
\end{defn}

\begin{defn}[Absolutely Maximally Contextual Correlations]
An $(n,m,o)$ correlation is called \emph{absolutely maximally contextual} if it is simultaneously maximally contextual and maximal marginal.
\label{def:amc}
\end{defn}

The no-signaling correlations for $(2,2,2)$-scenario form a convex polytope of dimension $8$ with $24$ vertices, such that $8$ are PR-boxes \cite{PRBox1994}. These $8$ PR-boxes constitute the AMCCs, as they correspond to $\mathrm{CF}=1$ and exhibit maximal marginals \cite{AMCC}. 
Considering the $(3,2,2)$-scenario, the AMCCs are GHZ correlations; for instance, the correlations obtained from the state $|\text{GHZ}\rangle = \frac{|000\rangle + |111\rangle}{\sqrt{2}}$ with measurements in the $X$ or $Y$ basis (e.g., $Y_1 = \sigma_x$, $Y_1^\prime = \sigma_y$, and similarly for the other two parties). The three-way correlations \cite{Barret2005} exhibit AMCCs. Along with these two examples, this empirical model with $8$ parameters provides a different class of AMCC \cite{AMCC}. Besides the AMCCs, a different class of correlations called non-AMCCs \cite{AMCC} is observed: maximal contextual correlations but not maximal marginals. We extend these results to the $(4,2,2)$-scenario in the next section, using the general construction \cite{AMCC}.

\section{The \texorpdfstring{$(4,2,2)$}{(4,2,2)} AMC and non-AMC Correlations}
\label{sec 02}
In this section, we introduce AMCC and non-AMCC for the $(4,2,2)$ scenario. Our characterization of these correlations is guided by the notion of AME states and the maximal contextual correlations $(\mathrm{CF}= 1)$. The clear distinction between AMCCs and AME states is that, while four-qubit AME states do not exist, four-party AMCCs do. The construction of AMCCs in the $(4,2,2)$-scenario involves two methods: the Constraint
Satisfaction Problem (CSP) method and the Parity Check Method. These are as follows.

\subsection{CSP Model}
\label{subsec 03}
The construction follows the structural properties of the correlations. The possibilistic form of the correlations, namely logical contextual models \cite{LogicalContextuality}, provides a clear understanding of this structural characterization. Given a measurement scenario $\langle \mathcal{Y}, \mathcal{M}, \mathcal{O} \rangle$, the \emph{possibilistic distribution} is defined over a Boolean semiring $\mathbb{B}=(\{0,1\}, \vee, \wedge, \neg)$, using the map $\bar{\mathcal{P}} : \mathcal{E} \to \mathbb{B}$ with $\sum_{y \in \mathcal{E}} \bar{\mathcal{P}}(y) = 1$ that defines an element of $D_{\mathbb{B}}(\mathcal{E})$. The compatible set of global sections with the support \cite{LogicalContextuality, AMCC} over the probabilistic empirical model $e = \{e_C\}_{C\in\mathcal{M}}$ is 
\begin{equation}
    S_e = \{\, s \in \mathcal{E}(\mathcal{Y}) \mid \forall C \in \mathcal{M},\; s|_C \in \operatorname{supp}(e_C) \,\}.
\end{equation}
The model is strongly contextual iff $S_e = \varnothing$, which implies there is no global distribution $d \in D_{\mathbb{B}}(\mathcal{Y})$ such that $d|_C \in \operatorname{supp}(e_C)$ for all $C \in \mathcal{M}$. The strongly contextual model implies maximal contextuality and vise versa.
The characterization follows the Constraint Satisfaction Problem (CSP), which reduces to the Boolean satisfiability problem
(SAT) for the dichotomic measurements case. In this respect, the empirical $e$ associates with a possibilistic model that consists of Boolean variables $\bar{\mathcal{Y}}$. Given a context $C \in \mathcal{M}$ and a section $s \in S_e(C)$ in the support, we can write the propositional formula
\begin{equation}
    B_{e} \;:=\; \bigwedge_{C \in \mathcal{M}} 
   \Bigl( \; \bigvee_{s \in S_{e}(C)} b_{s} \;\Bigr),
   \label{eq: Be}
\end{equation}
where each $b_s$ is a Boolean statement encoding assignments associated with the section $s$:
\begin{equation}
    b_{s} \;:=\;
\Bigl( \; \bigwedge_{\substack{Y \in C \\ s(Y) = 1}} Y \;\Bigr)
\;\wedge\;
\Bigl( \; \bigwedge_{\substack{Y \in C \\ s(Y) = 0}} \neg Y \;\Bigr).
\end{equation}
Thus $B_e$ is a conjunction over the contexts of the disjunctions of all local assignments in the support. If $B_e=0$, then the formula is unsatisfiable; equivalently, all propositions $B_i \;  \forall C\in \mathcal{M}$ are simultaneously unsatisfiable, i.e., $S_e = \varnothing$.

\subsection{Parity Check method}
\label{subsec 04}
This method provides Mermin's type of proof while being suitable for the construction of the symmetric AMCCs \cite{AMCC}. The method consists of set of $N$ (all possible contexts in $\mathcal{M}$, i.e., $N=2^n$) linear equations $\{L_i:i\in N\}$, namely \emph{parity equations}. Each equation $L_i$ has the form $\sum Y = \mathrm{P_i} \;(\mathrm{mod}\;2)$, where $Y\in\mathcal{Y}$ and the parity $\mathrm{P_i}\in\{0,1\}$. The $\mathrm{P_i}=0$ exhibits even parity, while $\mathrm{P_i}=1$ corresponds to odd parity. If $\sum_i^N L_i\neq\sum_i^NP_i$, then it leads to the same contradiction as $B_e=0$, which means $S_e = \varnothing$. Equivalently, all the parity equations are not simultaneously satisfiable by any possible assignments.

\subsection{Construction of AMCC and non-AMCC}
\label{subsec 05}
The first construction is followed by the parity check method, which produces symmetric AMCCs. The $(4,2,2)$-scenario has a measurement set $\mathcal{Y}$ in Eq.~(\ref{422 measurment set}). The corresponding measurement cover $\mathcal{M}$ contains $16$ contexts 
   \(\{
\{Y_1, Y_2, Y_3, Y_4\},
\{Y_1, Y_2, Y_3, Y_4^\prime\},
\{Y_1, Y_2, Y_3^\prime, Y_4\},
\{Y_1, Y_2,\\ Y_3^\prime, Y_4^\prime\},
\{Y_1, Y_2^\prime, Y_3, Y_4\},
\{Y_1, Y_2^\prime, Y_3, Y_4^\prime\},
\{Y_1, Y_2^\prime, Y_3^\prime, Y_4\},
\{\\Y_1,Y_2^\prime, Y_3^\prime, Y_4^\prime\},
\{Y_1^\prime, Y_2, Y_3, Y_4\},
\{Y_1^\prime, Y_2, Y_3, Y_4^\prime\},
\{Y_1^\prime, Y_2, Y_3^\prime,\\ Y_4\},
\{Y_1^\prime, Y_2, Y_3^\prime, Y_4^\prime\},
\{Y_1^\prime, Y_2^\prime, Y_3, Y_4\},
\{Y_1^\prime, Y_2^\prime, Y_3, Y_4^\prime\},
\{Y_1^\prime,\\ Y_2^\prime, Y_3^\prime, Y_4\},
\{Y_1^\prime, Y_2^\prime, Y_3^\prime, Y_4^\prime\}
\}\). The parity equations associated with it are:
\begin{equation}
\begin{aligned}
 Y_1 + Y_2 + Y_3 + Y_4 &= \mathrm{P}_1 \quad (\mathrm{mod}\;2),\;\\
Y_1 + Y_2 + Y_3 + Y_4^\prime &= \mathrm{P}_2 \quad (\mathrm{mod}\;2),\\
Y_1 + Y_2 + Y_3^\prime + Y_4 &= \mathrm{P}_3 \quad (\mathrm{mod}\;2),\;\\
Y_1 + Y_2 + Y_3^\prime + Y_4^\prime &= \mathrm{P}_4 \quad (\mathrm{mod}\;2),\\
 Y_1 + Y_2^\prime + Y_3 + Y_4 &= \mathrm{P}_5 \quad (\mathrm{mod}\;2),\;\\
Y_1 + Y_2^\prime + Y_3 + Y_4^\prime &= \mathrm{P}_6 \quad (\mathrm{mod}\;2),\\
  Y_1 + Y_2^\prime + Y_3^\prime + Y_4 &= \mathrm{P}_7 \quad (\mathrm{mod}\;2),\;\\
Y_1 + Y_2^\prime + Y_3^\prime + Y_4^\prime &= \mathrm{P}_8 \quad (\mathrm{mod}\;2),\\
Y_1^\prime + Y_2 + Y_3 + Y_4 &= \mathrm{P}_9 \quad (\mathrm{mod}\;2),\;\\
Y_1^\prime + Y_2 + Y_3 + Y_4^\prime &= \mathrm{P}_{10} \quad (\mathrm{mod}\;2),\\
Y_1^\prime + Y_2 + Y_3^\prime + Y_4 &= \mathrm{P}_{11} \quad (\mathrm{mod}\;2),\;\\
Y_1^\prime + Y_2 + Y_3^\prime + Y_4^\prime &= \mathrm{P}_{12} \quad (\mathrm{mod}\;2),\\
 Y_1^\prime + Y_2^\prime + Y_3 + Y_4 &= \mathrm{P}_{13} \quad (\mathrm{mod}\;2),\;\\
Y_1^\prime + Y_2^\prime + Y_3 + Y_4^\prime &= \mathrm{P}_{14} \quad (\mathrm{mod}\;2),\\
  Y_1^\prime + Y_2^\prime + Y_3^\prime + Y_4 &= \mathrm{P}_{15} \quad (\mathrm{mod}\;2),\;\\
Y_1^\prime + Y_2^\prime + Y_3^\prime + Y_4^\prime &= \mathrm{P}_{16} \quad (\mathrm{mod}\;2).
\end{aligned}
\label{eq:parity}
\end{equation}
Given the parity variable value $\mathrm{P}_i=0$ or $\mathrm{P}_i=1$, the corresponding equation will be reduced to the CSP model, where each proposition will take boolean statements $b_{s}$ from the set $\{(\neg Y_1 \wedge \neg Y_2 \wedge \neg Y_3 \wedge \neg Y_4),(\neg Y_1 \wedge \neg Y_2 \wedge Y_3 \wedge Y_4),(\neg Y_1 \wedge Y_2 \wedge \neg Y_3 \wedge Y_4), (\neg Y_1 \wedge Y_2 \wedge Y_3 \wedge \neg Y_4), (Y_1 \wedge \neg Y_2 \wedge \neg Y_3 \wedge Y_4), (Y_1 \wedge \neg Y_2 \wedge Y_3 \wedge \neg Y_4), (Y_1 \wedge Y_2 \wedge \neg Y_3 \wedge \neg Y_4), (Y_1 \wedge Y_2 \wedge Y_3 \wedge Y_4)\}$ or from the set $\{(\neg Y_1 \wedge \neg Y_2 \wedge \neg Y_3 \wedge Y_4), (\neg Y_1 \wedge \neg Y_2 \wedge Y_3 \wedge \neg Y_4), (\neg Y_1 \wedge Y_2 \wedge \neg Y_3 \wedge \neg Y_4), (\neg Y_1 \wedge Y_2 \wedge Y_3 \wedge Y_4), (Y_1 \wedge \neg Y_2 \wedge \neg Y_3 \wedge \neg Y_4), (Y_1 \wedge \neg Y_2 \wedge Y_3 \wedge Y_4), (Y_1 \wedge Y_2 \wedge \neg Y_3 \wedge Y_4), (Y_1 \wedge Y_2 \wedge Y_3 \wedge \neg Y_4)\}$ correspondingly. These sets constitute the variables associated with the equation; for instance, if we take the second parity equation, then the set of $b_s$ involves the variables $(Y_1,Y_2,Y_3,Y_4^\prime$ and so on. By construction, among all possible combinations of parity values $2^{16}$, the sets of parity equations corresponding to $65504$ combinations are simultaneously unsatisfiable for all $2^8$ global assignments over the measurement set $\mathcal{Y}$ defined in Eq.~(\ref{422 measurment set}). For instance, one of the unsatisfiable sets of parity equations corresponds to the parity conditions $\mathrm{P_1}=\mathrm{P_2}=\mathrm{P_3}=\mathrm{P_4}=\mathrm{P_5}=\mathrm{P_6}=\mathrm{P_7}=\mathrm{P_8}=\mathrm{P_9}=\mathrm{P_{10}}=\mathrm{P_{14}}=\mathrm{P_{15}}=\mathrm{P_{16}}=0$ and $\mathrm{P_{11}}=\mathrm{P_{12}}=\mathrm{P_{13}}=1$.
The associated CSP model construction follows from the fact that the last parity equation $Y_1^\prime + Y_2^\prime + Y_3^\prime + Y_4^\prime = 0 \quad (\mathrm{mod}\;2)$ is of even parity, so it takes the variables of that equation $(Y_1^\prime,Y_2^\prime,Y_3^\prime,Y_4^\prime)$ and the corresponding boolean statements associated with even parity can be written as   
\[
\begin{aligned}
B_{16} &= (\neg Y_1^\prime \wedge \neg Y_2^\prime \wedge \neg Y_3^\prime \wedge \neg Y_4^\prime)\vee(\neg Y_1^\prime \wedge \neg Y_2^\prime \wedge Y_3^\prime \wedge Y_4^\prime)\\
&\vee (\neg Y_1^\prime \wedge Y_2^\prime \wedge \neg Y_3^\prime \wedge Y_4^\prime) 
\vee (\neg Y_1^\prime \wedge Y_2^\prime \wedge Y_3^\prime \wedge \neg Y_4^\prime)\\
&\vee (Y_1^\prime \wedge \neg Y_2^\prime \wedge \neg Y_3^\prime \wedge Y_4^\prime)\vee (Y_1^\prime \wedge \neg Y_2^\prime \wedge Y_3^\prime \wedge \neg Y_4^\prime)\\ 
&\vee (Y_1^\prime \wedge Y_2^\prime \wedge \neg Y_3^\prime \wedge \neg Y_4^\prime)\vee (Y_1^\prime \wedge Y_2^\prime \wedge Y_3^\prime \wedge Y_4^\prime),
\end{aligned}
\]
similarly, the $B_15$ proposition corresponding to the 15\textsuperscript{th} parity equation $Y_1^\prime + Y_2^\prime + Y_3^\prime + Y_4 = 0 \quad (\mathrm{mod}\;2)$ consists of variables $(Y_1^\prime,Y_2^\prime,Y_3^\prime,Y_4)$ along with the boolean statement associated with the even parity set of $b_s$
\[
\begin{aligned}
B_{15} &= (\neg Y_1^\prime \wedge \neg Y_2^\prime \wedge \neg Y_3^\prime \wedge \neg Y_4)\vee(\neg Y_1^\prime \wedge \neg Y_2^\prime \wedge Y_3^\prime \wedge Y_4)\\
&\vee (\neg Y_1^\prime \wedge Y_2^\prime \wedge \neg Y_3^\prime \wedge Y_4) 
\vee (\neg Y_1^\prime \wedge Y_2^\prime \wedge Y_3^\prime \wedge \neg Y_4)\\
&\vee (Y_1^\prime \wedge \neg Y_2^\prime \wedge \neg Y_3^\prime \wedge Y_4)\vee (Y_1^\prime \wedge \neg Y_2^\prime \wedge Y_3^\prime \wedge \neg Y_4)\\ 
&\vee (Y_1^\prime \wedge Y_2^\prime \wedge \neg Y_3^\prime \wedge \neg Y_4)\vee (Y_1^\prime \wedge Y_2^\prime \wedge Y_3^\prime \wedge Y_4),
\end{aligned}
\]
the same pattern persists for the propositions $B_{1},B_{2},\cdots,B_{10},B_{14}$; In contrast, for propositions $B_{11},B_{12}$ and $B_{13}$, correspond to the odd parity.. For instance, the parity equation $Y_1^\prime + Y_2^\prime + Y_3 + Y_4 = 1 \quad (\mathrm{mod}\;2)$ takes the boolean statements $b_s$ from the odd parity set, which can be written as 
\[
\begin{aligned}
B_{13} &= (\neg Y_1^\prime \wedge \neg Y_2^\prime \wedge \neg Y_3 \wedge Y_4)\vee(\neg Y_1^\prime \wedge \neg Y_2^\prime \wedge Y_3 \wedge \neg Y_4)\\
&\vee (\neg Y_1^\prime \wedge Y_2^\prime \wedge \neg Y_3 \wedge \neg Y_4) 
\vee (\neg Y_1^\prime \wedge Y_2^\prime \wedge Y_3 \wedge Y_4)\\
&\vee (Y_1^\prime \wedge \neg Y_2^\prime \wedge \neg Y_3 \wedge \neg Y_4)\vee (Y_1^\prime \wedge \neg Y_2^\prime \wedge Y_3 \wedge Y_4)\\ 
&\vee (Y_1^\prime \wedge Y_2^\prime \wedge \neg Y_3 \wedge Y_4)\vee (Y_1^\prime \wedge Y_2^\prime \wedge Y_3 \wedge \neg Y_4).
\end{aligned}
\]
Considering the 15\textsuperscript{th} and 16\textsuperscript{th} parity equations $Y_1^\prime + Y_2^\prime + Y_3^\prime + Y_4 = 0 \quad (\mathrm{mod}\;2)$ and $ \; Y_1^\prime + Y_2^\prime + Y_3^\prime + Y_4^\prime = 0 \quad (\mathrm{mod}\;2)$, respectively, yields the condition $Y_4=Y_4^\prime$; however, taking the 13\textsuperscript{th} parity equation $Y_1^\prime + Y_2^\prime + Y_3 + Y_4 = 1 \quad (\mathrm{mod}\;2)$ with the 14\textsuperscript{th} parity equation $ \; Y_1^\prime + Y_2^\prime + Y_3 + Y_4^\prime = 0$ implies the condition $Y_4\neq Y_4^\prime$, which is a clear contradiction. This demonstrates that, for given assignments, the 14\textsuperscript{th}, 15\textsuperscript{th}, and 16\textsuperscript{th} parity equations are satisfiable, but the 13\textsuperscript{th} parity equation is unsatisfiable; consequently, the corresponding boolean proposition $B_{13}=0$ reduces to $B_e=0$. Hence $S_e = \varnothing$. The associated probabilistic model is therefore maximally contextual; together with this, the symmetric structure ensures maximal marginality; hence, the corresponding correlation is AMCC. The probability value corresponding to each boolean statement of a boolean proposition $B_i$ is $1/8$; others are assigned $0$.

Next, we construct the non-AMCC correlations using the constraint satisfaction problem (CSP) model. We begin by selecting Boolean propositions corresponding to the parity conditions $\mathrm{P_1}=\mathrm{P_2}=\mathrm{P_3}=\mathrm{P_4}=\mathrm{P_5}=\mathrm{P_6}=\mathrm{P_7}=\mathrm{P_8}=\mathrm{P_9}=\mathrm{P_{10}}=\mathrm{P_{14}}=\mathrm{P_{15}}=\mathrm{P_{16}}=0$ and $\mathrm{P_{11}}=\mathrm{P_{12}}=\mathrm{P_{13}}=1$, which were introduced above; however, other assignments may also be considered. Since each boolean proposition has $8$ boolean statements out of $16$ such choices. In principle, we can take other possible combinations by adding the remaining Boolean statements to each proposition. However, this yields $2^{128}-1$ possible combinations, making brute-force infeasible. To reduce the complexity, we restrict our analysis to the $9$ propositions corresponding to the contexts \(
\{Y_1, Y_2, Y_3, Y_4\},\{Y_1, Y_2, Y_3, Y_4^\prime\},
\{Y_1, Y_2, Y_3^\prime, Y_4^\prime\},
\{Y_1, Y_2^\prime,\\ Y_3, Y_4^\prime\},
\{Y_1, Y_2^\prime, Y_3^\prime, Y_4\},
\{Y_1^\prime, Y_2, Y_3, Y_4^\prime\},
\{Y_1^\prime, Y_2, Y_3^\prime, Y_4\},\\
\{Y_1^\prime, Y_2^\prime, Y_3, Y_4\}, \{Y_1^\prime, Y_2^\prime, Y_3, Y_4^\prime\}\). For each proposition, instead of selecting the remaining Boolean statements consecutively, we choose them randomly under controlled constraints. Specifically, we select the remaining $b_s$ such that $3$ additional Boolean statements are included for $B_1, B_7, B_{11}$ and $B_{14}$, $5$ for $B_{13}$, $4$ for $B_{6}$, $2$ for $B_4$ and $1$ for each $B_{2}$ and $B_{10}$. By construction, even for $4$ such combinations, there are $4$ sets of boolean propositions that are simultaneously unsatisfiable, which also satisfy the boolean no-signaling condition \cite{AMCC}. One of the four sets of unsatisfiable Boolean propositions is obtained by adding the following extra Boolean statements: $B_1$ includes $(Y_1\wedge \neg Y_2\wedge \neg Y_3 \wedge \neg Y_4),(Y_1\wedge Y_2\wedge Y_3 \wedge \neg Y_4),(Y_1\wedge \neg Y_2\wedge Y_3 \wedge Y_4)$; $B_2$ includes $(\neg Y_1\wedge Y_2\wedge Y_3 \wedge Y_4^\prime)$; $B_4$ includes $(Y_1\wedge \neg Y_2\wedge Y_3^\prime \wedge Y_4^\prime),(Y_1\wedge Y_2\wedge \neg Y_3^\prime \wedge Y_4^\prime)$; $B_6$ includes $(\neg Y_1\wedge \neg Y_2^\prime\wedge Y_3 \wedge \neg Y_4^\prime), (Y_1\wedge Y_2^\prime\wedge Y_3 \wedge \neg Y_4^\prime), (Y_1\wedge \neg Y_2^\prime\wedge \neg Y_3 \wedge \neg Y_4^\prime), (\neg Y_1\wedge Y_2^\prime\wedge \neg Y_3 \wedge \neg Y_4^\prime)$; $B_7$ includes $(\neg Y_1\wedge \neg Y_2^\prime\wedge Y_3^\prime \wedge \neg Y_4), (Y_1\wedge \neg Y_2^\prime\wedge \neg Y_3^\prime \wedge \neg Y_4), (Y_1\wedge \neg Y_2^\prime\wedge Y_3^\prime \wedge Y_4)$; $B_{10}$ includes $(Y_1^\prime \wedge \neg  Y_2\wedge \neg Y_3 \wedge \neg Y_4^\prime)$; $B_{11}$ includes $(\neg Y_1^\prime \wedge  Y_2\wedge \neg Y_3^\prime \wedge Y_4), (Y_1^\prime \wedge  Y_2\wedge \neg Y_3^\prime \wedge Y_4), (Y_1^\prime \wedge \neg  Y_2\wedge Y_3^\prime \wedge \neg Y_4)$; $B_{13}$ includes $(\neg Y_1^\prime\wedge Y_2^\prime\wedge \neg Y_3 \wedge Y_4), (Y_1^\prime\wedge \neg Y_2^\prime\wedge \neg Y_3 \wedge Y_4), (\neg Y_1^\prime\wedge \neg Y_2^\prime\wedge \neg Y_3 \wedge \neg Y_4), (Y_1^\prime\wedge Y_2^\prime\wedge \neg Y_3 \wedge \neg Y_4), (\neg Y_1^\prime\wedge Y_2^\prime\wedge Y_3 \wedge \neg Y_4)$ and $B_{14}$ includes $(Y_1^\prime\wedge \neg Y_2^\prime\wedge \neg Y_3 \wedge \neg Y_4^\prime), (\neg Y_1^\prime\wedge \neg Y_2^\prime\wedge \neg Y_3 \wedge Y_4^\prime), (Y_1^\prime\wedge Y_2^\prime\wedge Y_3 \wedge \neg Y_4^\prime)$. Out of all $16$, only the $9$ propositions are modified; the remaining $7$ propositions are unchanged. Thus, the updated $9$ propositions with the additional Boolean statements are:
\begin{widetext}
\begin{equation}
\begin{aligned}
B_{1} &= (\neg Y_1 \wedge \neg Y_2 \wedge \neg Y_3 \wedge \neg Y_4)\vee(\neg Y_1 \wedge \neg Y_2 \wedge Y_3 \wedge Y_4)\vee (\neg Y_1 \wedge Y_2 \wedge \neg Y_3 \wedge Y_4)\\ 
&\vee (\neg Y_1 \wedge Y_2 \wedge Y_3 \wedge \neg Y_4)\vee ( Y_1 \wedge \neg Y_2 \wedge \neg Y_3 \wedge \neg Y_4)
\vee (Y_1 \wedge \neg Y_2 \wedge \neg Y_3 \wedge Y_4)\\
&\vee (Y_1 \wedge \neg Y_2 \wedge Y_3 \wedge \neg Y_4) 
\vee (Y_1 \wedge \neg Y_2 \wedge Y_3 \wedge Y_4)\vee (Y_1 \wedge Y_2 \wedge \neg Y_3 \wedge \neg Y_4)\\
&\vee (Y_1 \wedge Y_2 \wedge Y_3 \wedge \neg Y_4)\vee (Y_1 \wedge Y_2 \wedge Y_3 \wedge Y_4),\\
B_{2} &= (\neg Y_1 \wedge \neg Y_2 \wedge \neg Y_3 \wedge \neg Y_4^\prime)\vee(\neg Y_1 \wedge \neg Y_2 \wedge Y_3 \wedge Y_4^\prime)\vee (\neg Y_1 \wedge Y_2 \wedge \neg Y_3 \wedge Y_4^\prime)\\ 
&\vee (\neg Y_1 \wedge Y_2 \wedge Y_3 \wedge \neg Y_4^\prime)\vee (\neg Y_1 \wedge Y_2 \wedge Y_3 \wedge Y_4^\prime)
\vee (Y_1 \wedge \neg Y_2 \wedge \neg Y_3 \wedge Y_4^\prime)\\
&\vee (Y_1 \wedge \neg Y_2 \wedge Y_3 \wedge \neg Y_4^\prime) 
\vee (Y_1 \wedge Y_2 \wedge \neg Y_3 \wedge \neg Y_4^\prime)\vee (Y_1 \wedge Y_2 \wedge Y_3 \wedge Y_4^\prime),\\
B_{4} &= (\neg Y_1 \wedge \neg Y_2 \wedge \neg Y_3^\prime \wedge \neg Y_4^\prime)\vee(\neg Y_1 \wedge \neg Y_2 \wedge Y_3^\prime \wedge Y_4^\prime)\vee(\neg Y_1 \wedge Y_2 \wedge \neg Y_3^\prime \wedge Y_4^\prime)\\
&\vee (\neg Y_1 \wedge Y_2 \wedge Y_3^\prime \wedge \neg Y_4^\prime) 
\vee (Y_1 \wedge \neg Y_2 \wedge \neg Y_3^\prime \wedge Y_4^\prime)\vee (Y_1 \wedge \neg Y_2 \wedge Y_3^\prime \wedge \neg Y_4^\prime)\\
&\vee (Y_1 \wedge \neg Y_2 \wedge Y_3^\prime \wedge Y_4^\prime)
\vee (Y_1 \wedge Y_2 \wedge \neg Y_3^\prime \wedge \neg Y_4^\prime) 
\vee (Y_1 \wedge Y_2 \wedge \neg Y_3^\prime \wedge Y_4^\prime)\\
&\vee (Y_1 \wedge Y_2 \wedge Y_3^\prime \wedge Y_4^\prime),\\
B_{6} &= (\neg Y_1 \wedge \neg Y_2^\prime \wedge \neg Y_3 \wedge \neg Y_4^\prime)\vee(\neg Y_1 \wedge \neg Y_2^\prime \wedge Y_3 \wedge \neg Y_4^\prime)\vee(\neg Y_1 \wedge \neg Y_2^\prime \wedge Y_3 \wedge Y_4^\prime)\\
&\vee(\neg Y_1 \wedge Y_2^\prime \wedge \neg Y_3 \wedge \neg Y_4^\prime)\vee (\neg Y_1 \wedge Y_2^\prime \wedge \neg Y_3 \wedge Y_4^\prime) 
\vee (\neg Y_1 \wedge Y_2^\prime \wedge Y_3 \wedge \neg Y_4^\prime)\\
&\vee(Y_1 \wedge \neg Y_2^\prime \wedge \neg Y_3 \wedge \neg Y_4^\prime)\vee (Y_1 \wedge \neg Y_2^\prime \wedge \neg Y_3 \wedge Y_4^\prime)\vee (Y_1 \wedge \neg Y_2^\prime \wedge Y_3 \wedge \neg Y_4^\prime)\\
&\vee (Y_1 \wedge Y_2^\prime \wedge \neg Y_3 \wedge \neg Y_4^\prime) 
\vee (Y_1 \wedge Y_2^\prime \wedge Y_3 \wedge \neg Y_4^\prime)\vee (Y_1 \wedge Y_2^\prime \wedge Y_3 \wedge Y_4^\prime),\\
B_{7} &= (\neg Y_1 \wedge \neg Y_2^\prime \wedge \neg Y_3^\prime \wedge \neg Y_4)\vee(\neg Y_1 \wedge \neg Y_2^\prime \wedge Y_3^\prime \wedge \neg Y_4)\vee(\neg Y_1 \wedge \neg Y_2^\prime \wedge Y_3^\prime \wedge Y_4)\\
&\vee (\neg Y_1 \wedge Y_2^\prime \wedge \neg Y_3^\prime \wedge Y_4) 
\vee (\neg Y_1 \wedge Y_2^\prime \wedge Y_3^\prime \wedge \neg Y_4)\vee(Y_1 \wedge \neg Y_2^\prime \wedge \neg Y_3^\prime \wedge \neg Y_4)\\
&\vee (Y_1 \wedge \neg Y_2^\prime \wedge \neg Y_3^\prime \wedge Y_4)
\vee (Y_1 \wedge \neg Y_2^\prime \wedge Y_3^\prime \wedge \neg Y_4) 
\vee \vee(Y_1 \wedge \neg Y_2^\prime \wedge Y_3^\prime \wedge Y_4)\\
&\vee(Y_1 \wedge Y_2^\prime \wedge \neg Y_3^\prime \wedge \neg Y_4)\vee (Y_1 \wedge Y_2^\prime \wedge Y_3^\prime \wedge Y_4),\\
B_{10} &= (\neg Y_1^\prime \wedge \neg Y_2 \wedge \neg Y_3 \wedge \neg Y_4^\prime)\vee(\neg Y_1^\prime \wedge \neg Y_2 \wedge Y_3 \wedge Y_4^\prime)\vee (\neg Y_1^\prime \wedge Y_2 \wedge \neg Y_3 \wedge Y_4^\prime)\\ 
&\vee(\neg Y_1^\prime \wedge Y_2 \wedge Y_3 \wedge \neg Y_4^\prime)\vee(Y_1^\prime \wedge \neg Y_2 \wedge \neg Y_3 \wedge \neg Y_4^\prime)\vee(Y_1^\prime \wedge \neg Y_2 \wedge \neg Y_3 \wedge Y_4^\prime)\\
&\vee (Y_1^\prime \wedge \neg Y_2 \wedge Y_3 \wedge \neg Y_4^\prime)
\vee (Y_1^\prime \wedge Y_2 \wedge \neg Y_3 \wedge \neg Y_4^\prime) 
\vee(Y_1^\prime \wedge Y_2 \wedge Y_3 \wedge Y_4^\prime),\\
B_{11} &= (\neg Y_1^\prime \wedge \neg Y_2 \wedge \neg Y_3^\prime \wedge Y_4)\vee (\neg Y_1^\prime \wedge \neg Y_2 \wedge Y_3^\prime \wedge \neg Y_4)\vee (\neg Y_1^\prime \wedge Y_2 \wedge \neg Y_3^\prime \wedge \neg Y_4)\\
&\vee(\neg Y_1^\prime \wedge Y_2 \wedge \neg Y_3^\prime \wedge Y_4)
\vee (\neg Y_1^\prime \wedge Y_2 \wedge Y_3^\prime \wedge Y_4) 
\vee (Y_1^\prime \wedge \neg Y_2 \wedge \neg Y_3^\prime \wedge \neg Y_4)\\
&\vee (Y_1^\prime \wedge \neg Y_2 \wedge Y_3^\prime \wedge \neg Y_4)
\vee (Y_1^\prime \wedge \neg Y_2 \wedge Y_3^\prime \wedge Y_4) 
\vee (Y_1^\prime \wedge Y_2 \wedge \neg Y_3^\prime \wedge \neg Y_4)\\
&\vee (Y_1^\prime \wedge Y_2 \wedge \neg Y_3^\prime \wedge Y_4)\vee(Y_1^\prime \wedge Y_2 \wedge Y_3^\prime \wedge \neg Y_4),\\
B_{13} &= (\neg Y_1^\prime \wedge \neg Y_2^\prime \wedge \neg Y_3 \wedge \neg Y_4)\vee (\neg Y_1^\prime \wedge \neg Y_2^\prime \wedge \neg Y_3 \wedge Y_4)\vee(\neg Y_1^\prime \wedge \neg Y_2^\prime \wedge Y_3 \wedge \neg Y_4)\\
&\vee (\neg Y_1^\prime \wedge Y_2^\prime \wedge \neg Y_3 \wedge \neg Y_4) 
\vee (\neg Y_1^\prime \wedge Y_2^\prime \wedge \neg Y_3 \wedge Y_4)
\vee (\neg Y_1^\prime \wedge Y_2^\prime \wedge Y_3 \wedge \neg Y_4)\\
&\vee (\neg Y_1^\prime \wedge Y_2^\prime \wedge Y_3 \wedge Y_4) 
\vee (Y_1^\prime \wedge \neg Y_2^\prime \wedge \neg Y_3 \wedge \neg Y_4)\vee (Y_1^\prime \wedge \neg Y_2^\prime \wedge \neg Y_3 \wedge Y_4)\\
&\vee(Y_1^\prime \wedge \neg Y_2^\prime \wedge Y_3 \wedge Y_4)\vee(Y_1^\prime \wedge Y_2^\prime \wedge \neg Y_3 \wedge \neg Y_4)\vee(Y_1^\prime \wedge Y_2^\prime \wedge \neg Y_3 \wedge Y_4)\\
&\vee(Y_1^\prime \wedge Y_2^\prime \wedge Y_3 \wedge \neg Y_4),\\
B_{14} &= (\neg Y_1^\prime \wedge \neg Y_2^\prime \wedge \neg Y_3 \wedge \neg Y_4^\prime)\vee(\neg Y_1^\prime \wedge \neg Y_2^\prime \wedge \neg Y_3 \wedge Y_4^\prime)\vee (\neg Y_1^\prime \wedge \neg Y_2^\prime \wedge Y_3 \wedge Y_4^\prime)\\ 
&\vee (\neg Y_1^\prime \wedge Y_2^\prime \wedge \neg Y_3 \wedge Y_4^\prime)\vee ( \neg Y_1^\prime \wedge Y_2^\prime \wedge Y_3 \wedge \neg Y_4^\prime)
\vee (Y_1^\prime \wedge \neg Y_2^\prime \wedge \neg Y_3 \wedge \neg Y_4^\prime)\\
&\vee (Y_1^\prime \wedge \neg Y_2^\prime \wedge \neg Y_3 \wedge Y_4^\prime) 
\vee (Y_1^\prime \wedge \neg Y_2^\prime \wedge Y_3 \wedge \neg Y_4^\prime)\vee (Y_1^\prime \wedge Y_2^\prime \wedge \neg Y_3 \wedge \neg Y_4^\prime)\\
&\vee(Y_1^\prime \wedge Y_2^\prime \wedge Y_3 \wedge \neg Y_4^\prime)\vee (Y_1^\prime \wedge Y_2^\prime \wedge Y_3 \wedge Y_4^\prime),\\
\end{aligned}
\end{equation}
\end{widetext}
The corresponding probabilistic model can be deduced using the zero constraints \cite{AMCC}. For the $(4,2,2)$ scenario, the probabilistic empirical model consists of $256$, while imposing the no-signaling and normalization constraints reduces the number of free parameters to $80$. Incorporating the zero entries from the possibilistic model and assigning them to the corresponding variables in the probabilistic model with $80$ parameters, the number of parameters is further reduced to $1$. As shown in Tables~(\ref{tab:nonAMCCa}) and Table~(\ref{tab:nonAMCCb}), which together form a single $16\times 16$ table representing the $(4,2,2))$-scenario. It represents a non-AMCC, such that the contextual fraction $\mathrm{CF}=1$ but not the maximal marginal for the bound $1/8< q <1/4$. However, it becomes AMCC under the condition $q=1/8$.
\begin{widetext}
\begin{table*}[t]
\centering
\scriptsize
\setlength{\tabcolsep}{12pt}
\renewcommand{\arraystretch}{1.80}

\begin{tabular}{|c|c| c| c| c| c| c| c| c|}
\hline
\textbf{Con.\& Sec.} &
\textbf{(0,0,0,0)} & \textbf{(0,0,0,1)} & \textbf{(0,0,1,0)} & \textbf{(0,0,1,1)} &
\textbf{(0,1,0,0)} & \textbf{(0,1,0,1)} & \textbf{(0,1,1,0)} & \textbf{(0,1,1,1)} \\
\hline

$(0,0,0,0)$ & $q$ & $0$ & $0$ & $1/4-q$ & $0$ & $1/4-q$ & $q$ & $0$\\ \hline
$(0,0,0,1)$ & $q$ & $0$ & $0$ & $1/4-q$ & $0$ & $1/4-q$ & $q$ & $0$\\ \hline
$(0,0,1,0)$ & $q$ & $0$ & $0$ & $1/4-q$ & $0$ & $1/4-q$ & $q$ & $0$\\ \hline
$(0,0,1,1)$ & $q$ & $0$ & $0$ & $1/4-q$ & $0$ & $1/4-q$ & $q$ & $0$\\ \hline
$(0,1,0,0)$ & $q$ & $0$ & $0$ & $1/4-q$ & $0$ & $1/4-q$ & $q$ & $0$\\ \hline
$(0,1,0,1)$ & $q$ & $0$ & $0$ & $1/4-q$ & $0$ & $1/4-q$ & $q$ & $0$\\ \hline
$(0,1,1,0)$ & $q$ & $0$ & $0$ & $1/4-q$ & $0$ & $1/4-q$ & $q$ & $0$\\ \hline
$(0,1,1,1)$ & $q$ & $0$ & $0$ & $1/4-q$ & $0$ & $1/4-q$ & $q$ & $0$\\ \hline

$(1,0,0,0)$ & $q$ & $0$ & $0$ & $1/4-q$ & $0$ & $1/4-q$ & $q$ & $0$\\ \hline
$(1,0,0,1)$ & $q$ & $0$ & $0$ & $1/4-q$ & $0$ & $1/4-q$ & $q$ & $0$\\ \hline
$(1,0,1,0)$ & $0$ & $1/4-q$ & $q$ & $0$ & $q$ & $0$ & $0$ & $1/4-q$\\ \hline
$(1,0,1,1)$ & $0$ & $1/4-q$ & $q$ & $0$ & $q$ & $0$ & $0$ & $1/4-q$\\ \hline
$(1,1,0,0)$ & $2q-1/4$ & $1/4-q$ & $1/4-q$ & $0$ & $1/4-q$ & $0$ & $2q-1/4$ & $1/4-q$\\ \hline
$(1,1,0,1)$ & $q$ & $0$ & $0$ & $1/4-q$ & $0$ & $1/4-q$ & $q$ & $0$\\ \hline
$(1,1,1,0)$ & $q$ & $0$ & $0$ & $1/4-q$ & $0$ & $1/4-q$ & $q$ & $0$\\ \hline
$(1,1,1,1)$ & $q$ & $0$ & $0$ & $1/4-q$ & $0$ & $1/4-q$ & $q$ & $0$\\ \hline

\end{tabular}

\caption{non-AMCC for $(4,2,2)$-scenario part (A).}
\label{tab:nonAMCCa}
\end{table*}
\end{widetext}

\begin{widetext}
\begin{table*}[t]
\centering
\scriptsize
\setlength{\tabcolsep}{15pt}
\renewcommand{\arraystretch}{1.80}

\begin{tabular}{|c| c| c| c| c| c| c| c|}
\hline
\textbf{(1,0,0,0)} & \textbf{(1,0,0,1)} & \textbf{(1,0,1,0)} & \textbf{(1,0,1,1)} &
\textbf{(1,1,0,0)} & \textbf{(1,1,0,1)} & \textbf{(1,1,1,0)} & \textbf{(1,1,1,1)}\\
\hline

$0$ & $1/4-q$ & $q$ & $0$ & $q$ & $0$ & $0$ & $1/4-q$\\ \hline
$0$ & $1/4-q$ & $q$ & $0$ & $q$ & $0$ & $0$ & $1/4-q$\\ \hline
$0$ & $1/4-q$ & $q$ & $0$ & $q$ & $0$ & $0$ & $1/4-q$\\ \hline
$0$ & $1/4-q$ & $q$ & $0$ & $q$ & $0$ & $0$ & $1/4-q$\\ \hline
$0$ & $1/4-q$ & $q$ & $0$ & $q$ & $0$ & $0$ & $1/4-q$\\ \hline
$0$ & $1/4-q$ & $q$ & $0$ & $q$ & $0$ & $0$ & $1/4-q$\\ \hline
$0$ & $1/4-q$ & $q$ & $0$ & $q$ & $0$ & $0$ & $1/4-q$\\ \hline
$0$ & $1/4-q$ & $q$ & $0$ & $q$ & $0$ & $0$ & $1/4-q$\\ \hline

$0$ & $1/4-q$ & $q$ & $0$ & $q$ & $0$ & $0$ & $1/4-q$\\ \hline
$0$ & $1/4-q$ & $q$ & $0$ & $q$ & $0$ & $0$ & $1/4-q$\\ \hline
$q$ & $0$ & $0$ & $1/4-q$ & $0$ & $1/4-q$ & $q$ & $0$\\ \hline
$q$ & $0$ & $0$ & $1/4-q$ & $0$ & $1/4-q$ & $q$ & $0$\\ \hline
$1/4-q$ & $0$ & $2q-1/4$ & $1/4-q$ & $2q-1/4$ & $1/4-q$ & $1/4-q$ & $0$\\ \hline
$0$ & $1/4-q$ & $q$ & $0$ & $q$ & $0$ & $0$ & $1/4-q$\\ \hline
$0$ & $1/4-q$ & $q$ & $0$ & $q$ & $0$ & $0$ & $1/4-q$\\ \hline
$0$ & $1/4-q$ & $q$ & $0$ & $q$ & $0$ & $0$ & $1/4-q$\\ \hline

\end{tabular}

\caption{non-AMCC for $(4,2,2)$-scenario part (B).}
\label{tab:nonAMCCb}
\end{table*}
\end{widetext}

\section{Conclusion}
\label{sec: 03}
Contextuality provides a fundamental framework for characterizing non-classical phenomena and has been extensively investigated in quantum computation. Since nonlocality arises as a particular instance of contextuality, understanding the structure of correlations in multipartite systems associated with contextual models remains an active and largely unresolved area of research. While maximal entanglement has been rigorously analyzed, the corresponding correlations in general multipartite scenarios are still insufficiently understood. To investigate the maximality of such correlations, we employ the sheaf-theoretic approach, in which the contextual fraction $\mathrm{CF}$ serves as a quantitative measure of contextuality, taking values in the interval $[0,1]$, where $0$ corresponds to noncontextual correlations and $1$ to fully contextual correlations. Within this framework, we introduce a distinguished class of correlations, termed absolutely maximally contextual correlations (AMCCs) \cite{AMCC}, inspired by and analogous to the notion of absolutely maximally entangled (AME) states. These AMCCs are characterized as maximally contextual correlations with maximal marginals. For example, in the bipartite setting, Bell states constitute the AME states, whereas Popescu–Rohrlich (PR) boxes realize the AMCCs in the $(2,2,2)$ scenario. Similarly, for three-qubit systems, GHZ states are AME, and the corresponding GHZ correlations form the AMCCs. Several additional families of AMCCs have also been identified in \cite{AMCC}. In the present work, we extend the analysis of AMCCs to the $(4,2,2)$ scenario. Notably, an absolutely maximally entangled state of four qubits does not exist, leading to a significant conceptual and structural divergence between the AME and AMCC. Our study further extends the previous results on non-AMCC correlations \cite{AMCC}, which are maximally contextual yet fail to exhibit maximal marginals. The construction of both AMCC and non-AMCC classes is carried out using parity-check formulations and constraint satisfaction problem (CSP)–based methods.

These maximal correlations play a central role in device-independent quantum information protocols and in the certification of intrinsic randomness. The class of correlations analyzed here is particularly well suited to these applications, as discussed in \cite{AMCC}.
\begin{acknowledgements}
The authors thank Aravinda S. and Rahul V. for fruitful discussions. 
\end{acknowledgements}

\newpage
\bibliography{00_draft}

\end{document}